\documentclass[a4paper]{article}

\usepackage[english]{babel}
\usepackage[utf8x]{inputenc}
\usepackage{amsmath, amssymb, amsfonts, graphicx}
\usepackage{mathtext, euscript}
\usepackage{hyperref}
\usepackage{xcolor}
\usepackage{authblk}

\title{Theory of Turbulence for \(d \rightarrow \infty\): Four-Loop Approximation of the Renormalization Group} 

\author[1,2]{Loran Ts. Adzhemyan\thanks{Email: \texttt{l.adzhemyan@spbu.ru}}}    
\author[3]{Yury Kirienko\thanks{Email: \texttt{yury.kirienko@gmail.com}}}    
\affil[1]{Saint Petersburg State University, 7/9 Universitetskaya nab., St. Petersburg, 199034, Russian Federation}     
\affil[2]{Bogolyubov Laboratory of Theoretical Physics, Joint Institute for Nuclear Research, 141980 Dubna, Russian Federation}
\affil[3]{secjur GmbH, Steinhöft 9, 20459 Hamburg}
\begin{document}
\maketitle

\begin{abstract}

Within the framework of the renormalization group approach in the stochastic model of fully developed turbulence, the $\beta$-function has been calculated in the fourth order of perturbation theory for high-dimensional spaces $d \rightarrow \infty$. The position of the fixed point of the renormalization group in the fourth order of the $\varepsilon$-expansion has been determined, and the index $\omega$, which defines the infrared stability of this point, has been calculated. We demonstrate the possibility of significantly reducing the number of Feynman diagrams through mutual cancellation. The results obtained allow us to find the four terms of the $\varepsilon$-expansion of the index $\omega$:
\begin{equation*}
  \omega = 2\,\varepsilon + \frac{2}{3}\,\varepsilon^2 + \frac{10}{9}\,\varepsilon^3 + \frac{56}{27}\,\varepsilon^4.
\end{equation*}

\end{abstract}

\section{Introduction}
Turbulence theory and second-order phase transitions share common features, such as universality in the critical region and the presence of scaling with non-trivial exponents. Both theories are described by certain quantum field models. The renormalization group method and the $\varepsilon$-expansion are used to justify critical scaling and calculate critical exponents. However, there are significant differences. In the theory of phase transitions, the parameter $\varepsilon=d_c-d$ is defined by the deviation of the space dimension $d$ from the critical dimension $d_c$, so that for $d>d_c$ the mean-field theory is valid. In turbulence theory, the parameter $\varepsilon$ is not related to the dimensionality of space, and the pair ${\varepsilon,d}$ can be considered as 2 independent parameters of the theory. To some extent, it is analogous to the pair ${\varepsilon, N}$ in the theory of phase transitions, where $N$ is the number of field components. In the latter case, in addition to the $\varepsilon$-expansion, the $1/N$-expansion is used to calculate critical exponents. 

Is it possible to implement an analogous approach in the form of a $1/d$-expansion in turbulence theory? Currently, such an expansion has been obtained in a simplified model of turbulent mixing of a passive scalar – the Kraichnan model \cite{Falk}. It has been shown that in the asymptotic $d\rightarrow\infty$, the Kolmogorov theory becomes valid, and the exponents of anomalous scaling -- deviations of critical exponents from Kolmogorov values -- have been calculated to order $O(1/d)$. There are indications that in the general theory of turbulence, $d=\infty$ may play the role of a "critical dimension", at which the Kolmogorov theory becomes valid \cite{Fourn}. In this case, the $1/d$-expansion would allow solving the problem of calculating anomalous exponents.

Currently, only a double ${\varepsilon,1/d}$ expansion in the leading order of $1/d$ and in the third order of $\varepsilon$ has been constructed \cite{dinf} (three-loop approximation), where the renormalization group functions in that work were calculated using the standard method through renormalization constants. In the paper \cite{Sazonov}, a more efficient method was proposed, in which the renormalization group functions themselves are the object of calculation based on Feynman diagrams, allowing the avoidance of calculating renormalization constants that contain poles in $\varepsilon$, which is especially convenient for numerical calculations. In the present work, we use the technique from \cite{Sazonov} to perform calculations in the four-loop approximation.

\section{Stochastic Model of Turbulence}
The motion of a viscous incompressible fluid is described by the Navier-Stokes equation
\begin{equation}
\label{NS}
\partial_t v_i=-\partial_i P-(v_j\partial_j)v_i+\nu_0 \partial^2 v_i+f_i\,,
\end{equation}
where $v_i$ is the velocity field, $P$ is the pressure, $\nu_0$ is the kinematic viscosity, and $f_i$ is the external force per unit mass.
The equation (\ref{NS}) is comple\-mented by the incompressibility condition 
\begin{equation}
\partial_i v_i = 0,
\end{equation}
which leads to the transversality of the velocity field and the force.
In the stochastic model of homogeneous isotropic turbulence, the force $f$ is considered random, simulating the pumping of energy into the system, compensating for viscous losses and ensuring the existence of a stationary state. It is usually assumed to have a simple Gaussian distribution with zero mean and correlator
\begin{equation}\label{ff}
\langle f_i(t_1,{\bf x}_1)f_j(t_2,{\bf x}_2)   \rangle\equiv D_{ij}^f(t_1-t_2,{\bf x}_1-{\bf x}_2)\,,
\end{equation}
\begin{equation}\label{D}
D_{ij}^f(t,{\bf k})=\delta (t)P_{ij}({\bf k})d_f(k)\,,
\end{equation}
where $P_{ij}({\bf k})\equiv \delta_{ij}-k_ik_j/k^2$ is the transverse projector. The spectral function of the energy pumping $d_f(k)$ is assumed to be localized in the region of the inverse sizes $L^{-1}$ of the largest vortices. If one is interested in the range of wave numbers $kL>>1$ (inertial interval and dissipation region), then for $d_f(k)$, one can use the model
\begin{equation}\label{df}
d_f(k)=A\delta ({\bf k}).
\end{equation}

A significant success in the study of developed turbulence was the pheno\-meno\-logical Kolmogorov's theory "K41", (see~\cite{K41}), many of whose predictions agree well with experiment. In particular, the famous law was predicted for the spectral energy density of turbulent fluctuations in the inertial interval
\begin{equation}\label{C}
\langle v_i({\bf k})v_i({-\bf k})   \rangle=C k^{-d-2/3}\,,
\end{equation}
where $d$ is the dimension of space. Experiments show that there are systematic deviations from the predictions of Kolmogorov's theory that increase with the order of the correlation functions (in particular, a slight shift in the exponent in (\ref{C})) -- so-called \textit{anomalous scaling}. The task of a consistent statistical theory can be seen as justifying Kolmogorov's hypotheses, determining the degree of their reliability and applicability, and also studying the phenomenon of anomalous scaling.

The most promising direction for studying the stochastic problem (\ref{NS})--(\ref{df}) is considered to be its reduction to a quantum field theory model. According to the fundamental theorem \cite{Martin}, the stochastic equation (\ref{NS})--(\ref{df})
is equivalent to a quantum field theory with a double set of transverse fields with the action given as
\begin{equation}\label{S0}
S_0 = v'D^f v'/2 + v' \left( -\partial_t v_i - (v\partial)v + \nu_0 \partial^2 v \right)\,.
\end{equation}
Here, all the necessary integrations and summations over indices are intended. The contribution of the pressure in (\ref{S0}) is omitted due to the transverseness of the auxiliary field $v'$. The action (\ref{S0}) has the typical appearance of a quantum field model with Yukawa-type interaction. For developed turbulence (at high Reynolds numbers), this interaction is very strong, and perturbation theory is not applicable.

A common approach to solving similar nonlinear problems can be described as follows. The original problem is extended to a non-physical domain of para\-meters. Such values of these para\-meters (critical values) are found at which the problem of finding the infrared asymptotics is solved exactly, and then a perturbation theory based on the deviation of the parameters from their critical values is constructed. These parameters may be naturally present in the model, or in some cases, they are introduced as additional parameters. A well-known example is Wilson's theory of critical phenomena, where the dimension of the space $d$ is considered as a continuously varying parameter of the theory. In the $\phi^4$ model the critical value is $d_c=4$, for $d>d_c$ the mean-field theory is valid, and for dimensions $d<d_c$ Wilson constructed an expansion by the formally small parameter $\varepsilon=4-d$. The same $\phi^4$ model also uses another approach - considering the number $N$ of field components as a parameter. As $N \rightarrow \infty$, the problem is reduced to the exactly solvable spherical model, and the theory for finite $N$ is constructed as a $1/N$ expansion.

An attempt was also made to apply the renormalization group method to the model (\ref{NS}). To do this, the delta function in the pumping function (\ref{df}) was first represented as the limit of a power function

\begin{equation}\label{delta}
\delta ({\bf k})=B \lim\limits_{\sigma\rightarrow +0}\frac{\sigma}{k^{d-\sigma}}.
\end{equation}

The presence of a dependence in (\ref{delta}) on $d$ leads to the absence of a critical dimension in Wilsonian turbulence theory. It is possible to construct a $\varepsilon$ expansion if we consider the parameter $\sigma$ in (\ref{delta}) as arbitrary. Its critical dimension turns out to be $\sigma_c=4$, and the formally small parameter $\varepsilon$ can be defined as $2\varepsilon=4-\sigma$ (the factor of two for convenience) with the physical value of $\varepsilon=2$. With this approach, it is possible to use the renormalization group and build expansions of physical quantities by the formally small parameter $\varepsilon$. For small $\varepsilon$, the fixed point of the renormalization group turns out to be infrared-stable, and for the energy spectrum at $\varepsilon=2$, the Kolmogorov expression (\ref{C}) is obtained, with the series in $\varepsilon$ for this quantity terminating at the first term.

Following the work of \cite{Sazonov}, we adopt the model for the function $d_f(k)$ from 
(\ref{D}) as
\begin{equation}\label{dfD}
d_f(k)=D_0k^{4-d-2\epsilon}\theta(k-m).
\end{equation}
The introduction of the factor $\theta(k-m)$ in (\ref{dfD}) helps to avoid infrared divergences in the perturbative diagrams.

The renormalization group approach in turbulence theory encounters some problems, in particular, the significant physical value of $\varepsilon$ raises doubts about the stability of the fixed point at the actual $\varepsilon=2$. As for the anomalous scaling, its origin may be related to the presence of composite operators with negative dimensions ("dangerous" operators) in the theory; however, at small $\varepsilon$ there are none, and at finite $\varepsilon$ they cannot be identified. The use of the renormalization group in combination with the analysis of dangerous operators to describe anomalous scaling has been successful in the simplified model of turbulent mixing of a passive scalar (the Kraichnan model), where the anomalous scaling exponents have been calculated up to the third order in $\varepsilon$ \cite{kabric}. An alternative way to find them in this model is to calculate them as a $1/d$ expansion. It turns out that at $d\rightarrow \infty$, the anomalous dimensions in the Kraichnan model go to zero and the Kolmogorov theory becomes valid (anomalous exponents of the order of $O(1/d)$ were computed in \cite{Falk}). In this sense, it can be said that in Kraichnan's theory, there exists a critical dimension of the space $d_c=\infty$ for which the Kolmogorov theory is valid. There are indications that this fact also holds for the full theory of turbulence. Taking advantage of this and learning to construct anomalous dimensions as a series in $O(1/d)$ would be tempting. In analogy to the $1/N$ expansion in the theory of critical phenomena, finding an analogue of the spherical model describing the asymptotic $d\rightarrow \infty$ would be necessary. Unfortunately, no such model is currently known. 

To have a constructive hint for constructing such a model, one could try to compute as many terms of the $\varepsilon$ expansion as possible at $d\rightarrow \infty$ and then sum the resulting series. In the paper \cite{dinf}, this program was run in the third order of perturbation theory (three-loop approximation), and the renormalization group functions in that paper were calculated in the standard way via renormalization constants. In the paper \cite{Sazonov} a more efficient method was proposed, in which the renormalization group functions are calculated directly from the Feynman diagrams, which allows to avoid the calculation of renormalization constants containing poles in $\varepsilon$, which is particularly convenient for numerical calculations. The results of the paper \cite{Sazonov} confirmed the answer of the paper \cite{dinf}. In the present work we use the technique of \cite{Sazonov} to perform calculations in the four-loop approximation.

\section{Renormalization Scheme. RG-functions expressed in terms of the renormalized Green functions}

The diagrams of perturbation theory defined by the action (\ref{S0}) contain ultraviolet (UV) divergences as $\varepsilon \rightarrow +0$. The invariance of the action under Galilean transformations results in only one divergent $1$-irreducible correlation function $\Gamma^{(0)}_{ij} = \langle v_i v_j' \rangle_{1-\text{ir}}$ (response function).
To cancel divergences in this function, only one counterterm of the type $v'\partial^2 v$ is required.
The renormalized action is given by
\begin{equation}\label{S}
S=v'D^f v'/2 + v'\left( -\partial_t v_i - (v\partial)v + \nu Z_\nu \partial^2 v \right)\,.
\end{equation}
It is derived from (\ref{S0}) by the multiplicative renormalization of the parameters:
\begin{equation}\label{ren}
D_0 = g_0\nu_0^3 = g \mu^{2\varepsilon}\nu^3\,,\quad \nu_0 = \nu Z_\nu\,,\quad g_0 = g \mu^{2\varepsilon} Z_g\,,\quad Z_g = Z_\nu^{-3}\,,
\end{equation}
where $\mu$ is the renormalization mass, $g$ is the dimensionless renormalized charge, and the renormalization of the fields is not required.

In the following, we use a renormalization scheme similar to renormalization at zero frequencies and moments, complemented by the additional condition $\mu = m$.
Let $\Gamma_{ij}(\mathbf{k}, \omega)$ be the $1$-irreducible response function calculated according to the action (\ref{S}) with the renormalization constant $Z_\nu = 1$. This function is proportional to the transverse projector:
\begin{equation}\label{Gamma}
\Gamma_{ij}(\mathbf{k},\omega) = P_{ij}(\mathbf{k})\Gamma(k,\omega)\,, \quad \Gamma(k,\omega) = \frac{\Gamma_{ii}(\mathbf{k},\omega)}{d-1}\,.
\end{equation}
Let us now define a normalized function that equals unity in the zero (loop-less) approximation:
\begin{equation}\label{normGamma}
\overline{\Gamma}(k,\omega) = \frac{\Gamma(k,\omega)}{-\nu k^2}\,.
\end{equation}
Then, for its renormalized version, we require the following normalization condition:
\begin{equation}\label{uslnorm}
\overline{\Gamma}^R|_{k=0,\omega=0,m=\mu} = 1\,,
\end{equation}
namely, all diagrammatic contributions to the renormalized function $\overline{\Gamma}^R$ must be canceled by the counterterms at the normalization point $k = 0$, $\omega = 0$, $m=\mu$. This condition defines the subtraction scheme and the form of the renormalization constant $Z_\nu$.

The Feynman diagrammatic technique corresponding to the model (\ref{S}), (\ref{D}), (\ref{dfD}), contains
the following propagators, given in the $(\mathbf{k}, t)$-representation by:
\begin{equation}\label{V}
\langle v_i(t_1)v_j(t_2) \rangle = \frac{d_f(k)}{2\nu k^2}\exp\big[-\nu k^2 |t_1 - t_2|\big] P_{ij}(\mathbf{k}) = -\hspace{-5pt}-\hspace{-5pt}-\hspace{-5pt}-\hspace{-5pt}-\hspace{-5pt}-\hspace{-5pt}-
\end{equation}
\begin{equation}\label{V'}
\langle v_i(t_1)v_j'(t_2) \rangle = \theta(t_1 - t_2)\exp\big[-\nu k^2 (t_1 - t_2)\big] P_{ij}(\mathbf{k}) = -\hspace{-5pt}-\hspace{-5pt}-\hspace{-5pt}-\hspace{-5pt}-\hspace{-5pt}|\hspace{-5pt}-
\end{equation}

The interaction in (\ref{S}) is represented by the triple vertex $-v'(v \partial)v = v'_j V_{jsl} v_s v_l$
with the vertex factor 
\begin{center}
\begin{eqnarray}
\nonumber
\label{Vert}
V_{jsl} = i k_s \delta_{jl} = -\hspace{-6pt}-\hspace{-6pt}-\hspace{-6pt}|\hspace{-6pt}-\hspace{-6pt}\Bigg\langle \\[-31pt]\bullet
\nonumber\\\,\,\,\,\,,
\end{eqnarray}
\end{center}
where $k_s$ is the momentum argument of the field $v'$.
The crossed endpoint in (\ref{Vert}) corresponds to the field $v'$, the endpoint marked by the bold dot corresponds to the field $v_s$ contracted with $i k_s$, and the plain line stands for the field $v_l$. 

We represent the perturbation series for the function $\overline{\Gamma}$ as
\begin{equation}
  \overline{\Gamma}(k, \omega, m, \mu) = 1 + \sum_{n \geq 1} u^n \mu^{2n\varepsilon} \sum_{i} \chi_{n}^{(i)}(k, \omega, m)\,,
  ~~~~u \equiv \frac{S_d g}{(2\pi)^d}\,,
\label{barGammaPert}
\end{equation}
where the $i$-summation runs over all $n$-loop diagrams of the function $\overline{\Gamma}$.
For convenience, we introduce a normalized charge $u$, in which $S_d$ is the surface area of the unit sphere in the $d$-dimensional space.

The renormalization constants $Z_\nu$ and $Z_g$ in our renormalization scheme (similar to the MS scheme) depend only on the
space dimension $d$ and parameter $\varepsilon$ and do not depend on the ratio $m / \mu$.
The equations of the renormalization group are obtained from the independence of the non-renormalized
Green functions on the parameter $\mu$ at fixed $\nu_0$ and $g_0$. The RG-equations look exactly the same as in the 
MS scheme \cite{Vasiliev}. In particular, the equation for the $1$-irreducible function $\Gamma^R$ is given by
\begin{equation}\label{RGeq}
(\mu \partial _\mu + \beta(g) \partial _g - \gamma_\nu(g) \,\nu \partial _\nu) \Gamma^R = 0\,,
\end{equation}
where
\begin{equation}\label{RGfunction}
\gamma_i(g) = \frac{-2\varepsilon g \partial _g \ln Z_i}{1 + g \partial _g \ln Z_g}\,, \quad \beta(g) = -g(2\varepsilon + \gamma _g) = -g(2\varepsilon - 3\gamma _\nu)\,.
\end{equation}
The latter equation in (\ref{RGfunction}) is a consequence of the relation between the renormalization constants $Z_g$ and $Z_\nu$ in (\ref{ren}).
The equations (\ref{RGfunction}) define the $\beta$ and $\gamma_\nu$ RG-functions in terms of the renormalization constants. These functions are finite
and do not contain poles in $\varepsilon$, due to the renormalizability of the theory. However, the required preliminary 
calculation of the singular in $\varepsilon$ renormalization constants is complicated, especially for the numerical evaluation of
$\beta$ and $\gamma_\nu$. Using equation (\ref{RGeq}), we express these RG-functions in terms of the renormalized
Green function $\Gamma^R$.

First of all, we derive the RG-equation for the normalized function $\overline{\Gamma}^R$.
Employing (\ref{normGamma}) and (\ref{uslnorm}) we find
\begin{equation}\label{RGeq1}
(\mu \partial_\mu + \beta(g) \partial_g - \gamma_\nu(g) \,\nu \partial_\nu)\overline{\Gamma}^R = \gamma_\nu \overline{\Gamma}^R\,.
\end{equation}
Considering this equation at the normalization point $k = 0$, $\omega = 0$, $m = \mu$ and taking into account that
\begin{equation}\label{NP}
\overline{\Gamma}^R|_{k=0,\omega=0}(m,\mu,\nu) = \overline{\Gamma}^R|_{k=0,\omega=0}(m/\mu)\,, \quad \partial_g \overline{\Gamma}^R|_{k=0,\omega=0,m=\mu} = 0,
\end{equation}
we obtain
\begin{equation}\label{gamma_nu}
\gamma_\nu(g) = -(m \partial_m \overline{\Gamma}^R)|_{k=0,\omega=0,m=\mu}\,.
\end{equation}
In (\ref{gamma_nu}), the RG-function $\gamma_\nu$ is written in terms of the renormalized function $\overline{\Gamma}^R$.
Usually, the computation of the renormalized functions involves the calculation of the divergent in $\varepsilon$
renormalization constants. To deal only with finite objects, we take into account the counterterms by the $R$-operation,
acting on the diagrams of the basic action, where $Z_\nu = 1$,
\begin{equation}\label{Rgamma}
\Gamma^R = R \Gamma = (1-K)R'\Gamma\,.
\end{equation}
Here, the $R'$-operation eliminates the divergences in the subgraphs of diagrams, and the operation $(1-K)$
removes the remaining superficial divergence. The $R'$-operation can be expressed as \cite{Zavialov}
\begin{equation}\label{Rgamma1}
R'\Gamma = \prod_j (1-K)_j \Gamma\,,
\end{equation}
where, for each diagram from $\Gamma$, the product runs over all its divergent subgraphs.
The renormalization operation (\ref{Rgamma}), (\ref{Rgamma1}) eliminates the divergences in the function
$\Gamma^R$ as a whole and separately in each diagram.

The formal UV-divergence index of the $1$-irreducible function $\langle vv' \rangle_{1-\text{ir}}$ is equal to $2$. This leads to the possible
counterterms of the $k^2$ and $i\omega$ types. However, as it is seen from (\ref{Vert}), the external leg $v'$ of 
this function is always multiplied by the external momentum $k$, therefore, only the divergence of the $k^2$-type
remains. In our renormalization scheme, this corresponds to the following subtraction operations
for the whole function $\overline{\Gamma}$ and
for the $1$-irreducible subgraphs of diagrams $\chi_n^{(i)}$ from (\ref{barGammaPert}), respectively:
\begin{eqnarray}\label{RR}
\nonumber
(1-K)\overline{\Gamma}(k,\omega,m,\mu) = \overline{\Gamma}(k,\omega,m,\mu) - \overline{\Gamma}|_{k=0,\omega=0,m=\mu}\,,\\
(1-K)\chi_j(k^2_j,\omega_j,m) = \chi_j(k^2_j,\omega_j,m) - \chi_j|_{k_j=0,\omega_j=0,m=\mu}\,,
\end{eqnarray}
where $k_j$ and $\omega_j$ are the momentum and frequency incoming to the subgraph $\chi_j$.

Note that the dimensionless counterterm $\overline{\Gamma}|_{k=0,\omega=0,m=\mu}$ does not depend on $m$, so
taking into account (\ref{gamma_nu}) we obtain
\begin{equation}\label{gamma_nu1}
\gamma_\nu(g) = -(m \partial_m R\overline{\Gamma})|_{k=0,\omega=0,m=\mu} = -(m \partial_m R'\overline{\Gamma})|_{k=0,\omega=0,m=\mu}\,.
\end{equation}
Substituting expansion (\ref{barGammaPert}) in (\ref{gamma_nu1}), we find
\begin{equation}\label{gamma_nu2}
\gamma_\nu(g) = \sum_{n \geq 1} u^n (\gamma_\nu)_n\,, \quad (\gamma_\nu)_n = -\mu^{2n\varepsilon} \sum_i (m \partial_m R' \chi_n^{(i)})|_{k=0,\omega=0,m=\mu}.
\end{equation}
The operations $m \partial_m$ and $R'$ in (\ref{gamma_nu2}) are commutative, so we can write
\begin{equation}\label{gamma_nu3}
\gamma_\nu(g) = \sum_{n \geq 1} u^n (\gamma_\nu)_n\,, \quad (\gamma_\nu)_n = -\mu^{2n\varepsilon} \sum_i (R' m \partial_m \chi_n^{(i)})|_{k=0,\omega=0,m=\mu}.
\end{equation}
In applying the $R'$ operation in the form (\ref{Rgamma1}) to (\ref{gamma_nu3}), it should be noted that in the terms where the derivative $\partial_m$ affects the variables of any essential subgraph, this subgraph becomes finite and the subtraction $(1-K)_j$ in (\ref{Rgamma1}) does not apply to it. The value $\gamma_\nu(g)$ in (\ref{gamma_nu3}) does not depend on $\mu$. This becomes explicit if, after differentiating $\partial_m$, we switch to integration over dimensionless momenta $q/\mu$, which allows, by introducing the operation
\begin{equation}\label{hat}
\hat{\partial}_m f(m) = \partial_m f(m)|_{m=1},
\end{equation}
to write (\ref{gamma_nu3}) as
\begin{equation}\label{gamma_nu4}
\gamma_\nu(g) = \sum_{n \geq 1} u^n (\gamma_\nu)_n\,, \quad (\gamma_\nu)_n = -\sum_i (R' \hat{\partial}_m \chi_n^{(i)})|_{k=0,\omega=0}.
\end{equation}
This is the main relation for the calculation of the RG-function $\gamma_\nu$. The $R'$ operation in (\ref{gamma_nu4}) acts on a quantity that does not depend on $m$, so in its defining relation (\ref{Rgamma1}), the subtraction operation can be defined as
\begin{eqnarray}\label{RR1}
(1-K)\chi_j(k^2_j,\omega_j) = \chi_j(k^2_j,\omega_j) - \chi_j|_{k_j=0,\omega_j=0}\,.
\end{eqnarray}

\section{Large $d$ Asymptotics}
Let us consider the diagrams in the momentum representation in the spherical coordinate system.
Then, the dimension of the space $d$ enters into the integration measure $\int_0^\infty\,dk\,k^{d-1}\,\int_0^{\pi}\,d\theta\,(\sin\theta)^{d-2}...$
and into the lines $vv$ of diagrams as $k^{2 - d - 2\varepsilon}$. The number of $vv$-lines in diagrams of the function $\Gamma$
coincides with the number of loops, consequently the "pure" integration momenta can always be associated with the $vv$-lines.
Then, the factor $\theta(k - m) k^{2 - d - 2\varepsilon}$ from (\ref{dfD}) changes $\int_0^\infty\,dk\,k^{d-1}$
to $\int_m^\infty\,dk\,k^{1 - 2\varepsilon}$, and the dependence on $d$ in the radial part disappears. When $d\rightarrow\infty$, the angular weight $(\sin\theta)^{d-2}$ has a sharp
maximum at $\theta = \pi / 2$. Since $\cos(\pi/2) = 0$, the inner products of the different internal integration momenta
vanish. Then, in the leading approximation at $d\rightarrow\infty$ one may consider the internal integration
momenta to be orthogonal to each other and to the external momentum $p$.
In this approximation, the integrands do not depend on angles, and the angular integrations give a factor $S_d$,
included in the definition of the charge $u$ (\ref{barGammaPert}). The latter charge is finite at the renormalization group fixed point.

Therefore, the main contribution to the Green functions at $d\rightarrow\infty$ is given by the diagrams without the inner products,
which drastically decreases the number of diagrams.

\section{Transition to Zero Frequency. R'-operation}
In the $(k,t)$-representation, the transition in (\ref{gamma_nu3}) to zero external frequency corresponds to integrating over all intermediate times. 
This integration with time exponential propagators (\ref{V}), (\ref{V'}) is easily performed. The integration scheme we used, as well as the implementation of the $R'$-operation, is illustrated in the following diagram:
\begin{equation} 
  \includegraphics[width=7cm]{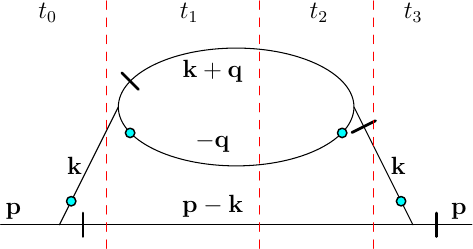}\label{fg1}
\end{equation}
Here, $k$ and $q$ are integration momenta, and $p$ is the external momentum of the 1-irreducible diagram. The flow of momenta is chosen so that the $vv$ lines correspond to simple momenta. The bold points correspond to momenta, and ignoring the scalar products, their contraction gives the factor $p^2 k^2$. According to the formula (\ref{gamma_nu3}), in the rest of the diagram, we can set $p=0$. The quantities $t_i$ denote the times of the corresponding vertices. The condition of zero external frequency in (\ref{gamma_nu3}) corresponds to integrating over all relative times of the diagram. Assuming $t_0=0$, we integrate over the remaining $t_i$. The presence of the $\theta$-function in the propagator (\ref{V'}) means that the integration region satisfies the inequalities:
\begin{equation} \label{usl}
t_1 \leq t_2 \leq t_3, \qquad t_3 \geq 0.
\end{equation}
Assuming the integration region in the diagram (\ref{fg1}) is chosen such that $t_1 \geq 0$, then all vertices are ordered by time, and the integrals over the relative times $\tau_1 = t_1 - t_0, \, \tau_2 = t_2 - t_1, \, \tau_3 = t_3 - t_2$ are taken within $0 \leq \tau_i < \infty$. The propagators (\ref{V}), (\ref{V'}) have an exponential dependence on time. Writing the exponent corresponding to the lower propagator as $ \exp\left[-E_{p-k}(t_3-t_0)\right] = \exp\left[E_{p-k}(\tau_1 + \tau_2 + \tau_3)\right]$, where $E_k = \nu k^2$ is the "energy" with momentum $k$, we achieve factorization of the integrals, each giving a denominator factor equal to the sum of the energies at the sections marked by dashed lines in the diagram:
\begin{equation}\label{E}
\frac{1}{\left(E_k + E_{p-k}\right) \left(E_{k+q} + E_{-q} + E_{p-k}\right) \left(E_k + E_{p-k}\right)}.
\end{equation}
For the contribution of the time version (\ref{fg1}), we obtain
\begin{equation}\label{Fg1}
\delta \chi = -\frac{\nu^3}{8} \int_m^\infty dk \int_m^\infty dq \frac{k^{1-2\varepsilon} q^{1-2\varepsilon} k^2}{\left(E_k + E_{-k}\right)^2 \left(E_{k+q} + E_{-q} + E_{-k}\right)}.
\end{equation}

To account for other integration regions, we need to consider the sum of "time versions" – all possible time orderings of the diagrams that are consistent with the conditions (\ref{usl}). In this case, there are two more time versions:
\begin{equation}
  \includegraphics[width=6.2cm]{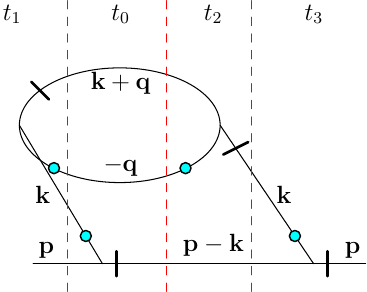} 
  \includegraphics[width=6.2cm]{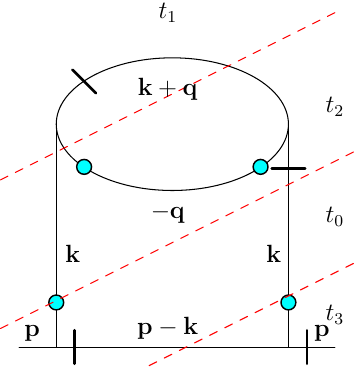 }\label{fg2}
\end{equation}
Dashed lines (sections) are drawn between each pair of nearest vertices, and the integration result is written in a form similar to (\ref{E}).

\subsection{ Implementation of the $R'$-Operation on a Diagram}
Regarding the implementation of the $R'$-operation on the diagram (\ref{fg1}). It contains a one-loop logarithmically divergent $vv'$ subgraph with vertices $t_1$ and $t_2$. According to (\ref{RR1}), the $R'$-operation subtracts from it the value at zero incoming frequency and momentum. In terms of the expression integrated over time, this can be done as follows. Introduce a parameter $a$ into the integrand such that the value $a=1$ gives the original diagram expression, and the value $a=0$ corresponds to zero values of the incoming frequency and momentum in the subgraph. The action of the $R'$ operation will then correspond to the difference of the obtained expressions. The parameter $a$ is introduced into the diagram as follows. In the factor $\left(E_{k+q}+E_{-q}+E_{p-k}\right)$, corresponding to the section in (\ref{E}) passing through the subgraph, place factors $a$ before the energy $E_{p-k}$, which does not belong to the subgraph, and before the incoming momentum in the subgraph energy $E_{k+q}$. As a result, we get $\left(E_{ak+q}+E_{-q}+aE_{p-k}\right)$. The necessary condition for $a=1$ is obviously satisfied. For $a=0$, the removal of the term $E_{p-k}$ corresponds to zero incoming frequency in the subgraph, and the factor $a=0$ in $E_{ak+q}$ eliminates the dependence on the incoming momentum in the subgraph.

Introducing the parameter $a$ into (\ref{Fg1}), taking into account that $E_k=\nu k^2$ and neglecting scalar products, we get
\begin{equation}\label{Fg1a}
\delta \chi_a = -\frac{1}{8} \int_m^\infty dk \int_m^\infty dq \frac{k^{3-2\varepsilon} q^{1-2\varepsilon}}{4k^4 \left(q^2 + a^2k^2 + q^2 + ak^2\right)}.
\end{equation}
Now consider that the action of the $R'$-operator in (\ref{RR1}) is preceded by differen\-tia\-tion $\hat{\partial}_m$.
The action of this operator gives the sum of contributions $\hat{\partial}_m = \sum_i \hat{\partial}_m^{(i)}$, where the limits of each integration variable $k_i$ are differentiated.
For the time version (\ref{fg1}), this gives 
\begin{equation}\label{Fg1aR}
R'\hat{\partial}_m \delta \chi = R'(\hat{\partial}_m^{(k)} + \hat{\partial}_m^{(q)}) \delta \chi.
\end{equation}
The operation $\hat{\partial}_m^{(q)}$ removes the subgraph divergence, so, considering (\ref{Fg1a}) with $a=1$, we have
\begin{equation}\label{q}
R'\hat{\partial}_m^{(q)} \delta \chi = \hat{\partial}_m^{(q)} \delta \chi = -\frac{1}{64} \int_1^\infty dk \frac{1}{k^{1+2\varepsilon} (k^2 + 1)}.
\end{equation}

To compute the second term in (\ref{Fg1aR}), we apply the operation $\hat\partial_m^{(k)}$ to the right-hand side of (\ref{Fg1a}) and take the difference of the values at $a=1$ and $a=0$:
\begin{equation}\label{k}
R'\hat\partial_m^{(k)} \delta \chi = -\frac{1}{64} \int_1^\infty dq \, q^{1-2\varepsilon} \left( \frac{1}{(q^2+1)} - \frac{1}{q^2} \right) = \frac{1}{64} \int_1^\infty dq \, \frac{1}{q^{1+2\varepsilon}(q^2+1)}.
\end{equation}
The obtained expressions (\ref{q}) and (\ref{k}) are finite at $\varepsilon=0$, and their $\varepsilon$-expansion can be found by expanding the integrands in a series.

In some time versions, the subgraph divergence may disappear if the number of sections through the subgraph in these versions exceeds the number of sections in the isolated subgraph. This is the case for the time version on the left part of (\ref{fg2}). In this case, the $R'$-operation is not needed. For the time version on the right part of (\ref{fg2}), the parameter $a$ is placed in the subgraph section factor according to the formulated rule: $\left(E_{ak+q} + E_{-q} + aE_{k}\right)$.

The generalization of the formulated rules allows us to compute $\gamma_{\nu}$ from (\ref{gamma_nu4}). The computation procedure was fully automated. It includes the construction of diagrams, the selection of significant ones (not containing scalar products), integration over time, and obtaining integrands for the contributions in (\ref{gamma_nu4}).

\section{Results}
The selection of significant diagrams showed that in the one-loop approximation, there is 1 significant diagram out of 4, in the two-loop approximation, there are 6 diagrams out of 120, in the three-loop approximation, there are 83 diagrams out of 4080, and in the four-loop approximation, there are 1692 out of 417872. In each approximation, the calculation was performed with consistent accuracy in the $\varepsilon$-expansion. Up to the three-loop approximation, the diagrams were calculated analytically, and in the four-loop approximation, they were calculated numerically. As a result, the following expression for $\gamma_\nu$ was obtained:
\begin{equation}
\label{otvet}
  \gamma_\nu=\frac{u}{4}+\frac{u^2}{32}\left(1-2\,\ln 2 \,\varepsilon+\frac{\pi^2}{6}\varepsilon^2\right)+\frac{u^3}{512}\left(7+6\,\ln 2 +a_3\,\varepsilon\right)+u^4a_4,
\end{equation}
where
\begin{equation}
\label{a3}
  a_3=-\frac{\pi^2}{2}+8-45\,\ln3+24\ln2-18(\ln2)^2-9\,\text{dilog} (3/4), 
\end{equation}
\begin{equation}
\label{a4a}
  a_4=0.02581007.
\end{equation}

Switching to the $\beta$ function (\ref{RGfunction}) for the charge $u$, substituting (\ref{otvet}) into it and solving the equation $\beta(u_*) = 0$,
we find the value of the charge at the fixed point in terms of the $\varepsilon$-expansion:
\begin{equation}
\label{ustar}
  u_* = \frac{8}{3}\,\varepsilon - \frac{8}{9}\,\varepsilon^2 - \frac{4 }{9}(1 - 2 \ln 2)\,\varepsilon^3 + \frac{4}{81}\left(25-18\,\ln2-3\,\pi^2-12a_3-72a_4\right)\varepsilon^4.
\end{equation}
The form of the renormalization group function $\gamma_\nu$, the associated $\beta$-function $\beta(u)=-u(\varepsilon-3\gamma _\nu)$, and the value of the charge at the fixed point $u_*$ depend on the chosen renormalization scheme. An objective characteristic is the value of the critical index $\omega=\partial_u \beta|_{u=u_*}$, which determines the IR stability of the fixed point by the condition $\omega>0$. The relations (\ref{otvet}) and (\ref{ustar}) allow us to find the four terms of the $\varepsilon$-expansion of the index $\omega$:
\begin{equation}
\label{omega}
  \omega = 2\,\varepsilon + \frac{2}{3}\,\varepsilon^2 + \frac{10}{9}\,\varepsilon^3 + \frac{1}{9}\left(\pi^2+\frac{46}{3}+2a_3+18a_4\right)\,\varepsilon^4.
\end{equation}
Substituting the expressions (\ref{a3}) and (\ref{a4a}) into (\ref{omega}) gives for the coefficient of $\varepsilon^4$ the value $2.07407470$, which very accurately matches the proper fraction $\frac{56}{27}=2.074074074...$.
This allows us to write
\begin{equation}\label{omega1}
  \omega = 2\,\varepsilon + \frac{2}{3}\,\varepsilon^2 + \frac{10}{9}\,\varepsilon^3 + \frac{56}{27}\,\varepsilon^4.
\end{equation}
Using the expression (\ref{omega1}), we can reconstruct the analytical result for the coefficient $a_4$ by requiring the elimination of irrational contributions:
\begin{equation}\label{a4}
 a_4 = \frac{9}{2048}\left(1 + 5\,\ln3 - \frac{8}{3}\ln2 + (\ln2)^2 + \text{dilog}(3/4)\right).
\end{equation}

\section{Conclusions}
We have carried out the calculations of the anomalous dimension $\gamma_\nu$ and exponent $\omega$ in the four-loop approximation in the model of fully developed turbulence in large dimensions of the space, using the method of determining RG-functions without renormalization constants. Our calculations demonstrated the efficiency of the applied method. The main advantage is that for the computation of the $n$-loop results, one needs to evaluate a set of $(n-1)$-dimensional integrals, free from singularities in $\varepsilon$. The computation procedure can be relatively easily automated. The main difficulty in the calculation of high orders of perturbation theory lies in the rapidly growing number of diagrams with vector fields and the triple interaction vertex. In the considered limit $d\rightarrow\infty$, the number of diagrams is notably decreased. Thus, in four loops, only $1692$ diagrams remain from a total of $417872$.

The significant decrease in the number of diagrams may indicate that the $\varepsilon$-expansion of the function $\gamma_\nu(\varepsilon)$ has a finite radius of convergence in the limit $d\rightarrow\infty$, in contrast to the general case, where one observes a factorial growth of the corresponding series coefficients. A similar situation occurs in the theory of phase transitions with the anomalous dimensions $\gamma(\varepsilon, N)$. Here, $\varepsilon = 4 - d$, $d$ is the dimension of the space, and $N$ is the number of field components. At fixed $N$, the coefficients of the $\varepsilon$-expansion grow factorially, but the coefficients of the $(1/N)$-expansion have a finite radius of convergence, as do the series in $\varepsilon$ \cite{Vasiliev}. Currently, in the theory of turbulence, only the third order in $\varepsilon$ at $d\rightarrow\infty$ of the double $(\varepsilon, 1/d)$-expansion is known. However, the results obtained in the current work and in \cite{dinf} have revealed some simplicity of this expansion. It is seen from equations (\ref{ustar}) and (\ref{omega}) that the irrational contribution to $u_*$ (the charge at the fixed point) with $\ln 2$ disappears in the physical quantity of the index $\omega$. The contributions of some particular diagrams proportional to $\ln^2 2$, $\pi^2$, and dilog(3/2), typical for the series of critical exponents in models of critical dynamics, are all canceled in the total sum of the diagram. The coefficients of (\ref{omega}) are rational numbers. This gives hope to summarize the series and find the function $\gamma_\nu(\varepsilon)$ at $d\rightarrow\infty$ without using the $\varepsilon$-expansion.

It should be noted that the approximation of this function proposed in \cite{dinf} predicted the coefficient of $\epsilon^4$ in (\ref{omega1}) to be 50/27, which slightly differs from our found coefficient of 56/27.

Our results of four-loop diagram calculations confirm the interesting effect of the "reduction" of diagrams discovered in \cite{Kirienko}, where certain groups of diagrams sum to zero. As shown in \cite{Kirienko}, the complete result in the two-loop approximation is given by the sum of 4 diagrams out of 6, and in the three-loop approximation by the sum of 9 diagrams out of 83. The 1692 four-loop diagrams we calculated also demonstrate such a reduction. Unfortunately, we have not yet been able to determine the minimal set of diagrams whose sum gives the result. Certain groups of mutually canceling diagrams have been identified, but they contain overlapping diagrams, preventing simultaneous elimination. The search for a minimal set by direct enumeration is very difficult due to the large number of diagrams. We hope to solve this problem in the future. Identifying the mechanism of reduction may bring us closer to solving the problem of summing the series.

\bibliographystyle{unsrt}
\bibliography{inf_turbulence.bib}

\begin{thebibliography}{10}

\bibitem{Falk}
M.~Chertkov, G.~Falkovich, I.~Kolokolov, and V.~Lebedev.
\newblock Normal and anomalous scaling of the fourth-order correlation function of a randomly advected passive scalar.
\newblock {\em Phys. Rev. E}, 52:4924--4941, Nov 1995.

\bibitem{Fourn}
J.-D Fournier, U.~Frisch, and H.A Rose.
\newblock Infinite-dimensional turbulence.
\newblock {\em J. Phys. A}, 11:187--198, 1978.

\bibitem{dinf}
L~Ts Adzhemyan, N~V Antonov, P~B Gol'din, T~L Kim, and M~V Kompaniets.
\newblock Renormalization group in the infinite-dimensional turbulence: third-order results.
\newblock {\em Journal of Physics A: Mathematical and Theoretical}, 41(49):495002, oct 2008.

\bibitem{Sazonov}
L~Ts Adzhemyan, TL~Kim, MV~Kompaniets, and VK~Sazonov.
\newblock Renormalization group in the infinite-dimensional turbulence: determination of the rg-functions without renormalization constants.
\newblock {\em Nanosystems: Physics, Chemistry, Mathematics}, 6(4):461--469, 2015.

\bibitem{K41}
Andrey~Nikolaevich Kolmogorov.
\newblock Dissipation of energy in the locally isotropic turbulence.
\newblock {\em Doklady Akademii Nauk SSSR}, 32(1):16--18, 1941.

\bibitem{Martin}
P.~C. Martin, E.~D. Siggia, and H.~A. Rose.
\newblock Statistical dynamics of classical systems.
\newblock {\em Phys. Rev. A}, 8:423--437, Jul 1973.

\bibitem{kabric}
L.~Ts. Adzhemyan, N.~V. Antonov, V.~A. Barinov, Yu.~S. Kabrits, and A.~N. Vasil'ev.
\newblock Calculation of the anomalous exponents in the rapid-change model of passive scalar advection to order ${\ensuremath{\varepsilon}}^{3}$.
\newblock {\em Phys. Rev. E}, 64:056306, Oct 2001.

\bibitem{Vasiliev}
A.~N. Vasil'ev.
\newblock {\em The Field Theoretic Renormalization Group in Critical Behavior Theory and Stochastic Dynamics}.
\newblock Routledge Chapman \& Hall, 2004.

\bibitem{Zavialov}
O.I. Zavialov.
\newblock {\em Renormalized Quantum Field Theory}.
\newblock Mathematics and its Applications. Springer Netherlands, 1990.

\bibitem{Kirienko}
Yury~V Kirienko and Tatyana~L Kim.
\newblock The stochastic model of turbulence: Simplification of the diagram technique in high dimensions.
\newblock 2016.

\end{thebibliography}

\end{document}